\documentclass[twocolumn,prl,showpacs]{revtex4}

\usepackage{graphicx}
\usepackage{rotating}
\usepackage{amsmath}
\usepackage{amsfonts}
\usepackage{amssymb}
\usepackage{enumerate}
\usepackage{longtable}
\usepackage{dcolumn}
\usepackage{bm}

\begin{document}

\title{Keeping a Quantum Bit Alive by Optimized $\pi$-Pulse
  Sequences}

\author{G\"otz S. Uhrig}
\email{goetz.uhrig@uni-dortmund.de}
\affiliation{Lehrstuhl f\"{u}r Theoretische Physik I,
 Universit\"{a}t Dortmund,
 Otto-Hahn Stra\ss{}e 4, 44221 Dortmund, Germany}

\date{\rm\today}

\begin{abstract}
A general strategy to maintain the coherence of a quantum bit
is proposed. The analytical result is derived rigorously
including all memory and back-action effects. 
It is based on an optimized $\pi$-pulse sequence
for dynamic decoupling extending the Carr-Purcell-Meiboom-Gill
(CPMG) cycle. The optimized sequence is very 
efficient, in particular for strong couplings to the environment.
\end{abstract}



\maketitle

Quantum information processing is a very promising and very challenging 
concept \cite{zolle05}. The basic feature which makes quantum information
conceptually more powerful than classical information is the quantum 
mechanical superposition principle. It allows for the parallel
processing of many classical registers -- the so-called quantum parallelism.
The single quantum bit (qubit) is a two-level system which we may
identify with a $S=1/2$ system with states $\downarrow$ and $\uparrow$.
Henceforth, we will use this spin language to characterize the qubit.
In order for this idea to work the qubit has to maintain its quantum state 
not only with respect to the state $\uparrow$ or $\downarrow$ but also
with respect to its relative phase. Unavoidable couplings between
the qubit and the environment spoil the quantum state: the qubit
loses its coherence. This decoherence is one of the most serious obstacles
on the way towards applications.
Hence finding strategies to suppress decoherence is a crucial
field of research \cite{zolle05}.

Dynamic decoupling \cite{viola98,ban98,facch05,cappe06} is one means to fight 
decoherence. The idea comes from spin-echo pulses in NMR
\cite{haebe76} where a large ensemble of spins is considered.
Static, but non-uniform couplings can be compensated perfectly
by a single $\pi$-pulse in the middle of the elapsing time interval.
The detrimental effect of more complicated perturbations like dynamic 
interactions with the environment can be suppressed by periodic
$\pi$-pulses or by periodic Carr-Purcell cycles of two $\pi$-pulses
each \cite{haebe76}.

The aim of this Letter is to show that an optimized sequence of
$\pi$-pulses suppresses the decoherence even more efficiently than
the so far known sequences of equidistant pulses
\cite{viola98,ban98,facch05,cappe06}.
The proposed scheme extends the known Carr-Purcell-Meiboom-Gill (CPMG)
cycle \cite{haebe76,witze06}.
In particular for a strong coupling to the environment, the optimized
sequence achieves a much better suppression for the same number of pulses.
So the optimized sequences will help to come closer to the realization
of quantum information devices.

We consider a fully quantum mechanical model
\begin{equation}
\label{eq:hamilton}
H=\sum_i\omega_i b_i^{\dagger}b_i^{\phantom\dagger}+\frac{1}{2}
\sigma_z\sum_i \lambda_i(b_i^{\dagger} + b_i^{\phantom\dagger}) + E
\end{equation}
consisting of a single qubit interpreted as spin $S=1/2$,
whose operators are the Pauli matrices $\sigma_x, \sigma_y$, and
$\sigma_z$. The environment is represented by a bosonic bath with annihilation 
(creation) operators $b_i^{(\dagger)}$. 
The constant $E$ sets the energy offset.
The relevant bath properties
are given by the spectral density \cite{legge87,weiss99}
\begin{subequations}
\label{specfunc}
\begin{eqnarray}
J(\omega) &=& \sum_i |\lambda_i|^2 \delta(\omega-\omega_i)\\
\label{eq:ohmic}
 &=& 2\alpha \omega \Theta(\omega_\text{D}-\omega)\ ,
\end{eqnarray}
\end{subequations}
where we have chosen the standard ohmic bath with linearly
rising density in (\ref{eq:ohmic}); $\alpha$ is the dimensionless
parameter controling the coupling between qubit and bath.
But our scheme proposed below can be applied to any spectral density 
$J(\omega)$.
The high-energy cutoff $\omega_\text{D}$ is chosen 
as in a Debye model for phonons, but other choices are equally possible.
Note that the correlation time $t_\text{C}$ of the bath  is
set by $1/\omega_\text{D}$.

The model (\ref{eq:hamilton}) can be easily diagonalized by the unitary
transformation $U=\exp(\sigma_z K)$ where 
$K=\sum_i (\lambda_i/(2\omega_i))(b_i^{\dagger} - b_i^{\phantom\dagger})$
is antihermitean. The resulting effective Hamiltonian
\footnote{The operator $A^\text{ef}$  is generally given by 
$A^\text{ef}= UAU^\dagger$.}
$H^\text{ef}=\sum_i\omega_i b_i^{\dagger}b_i^{\phantom\dagger}+\Delta E$
is manifestly diagonal with 
$\Delta E = E-\int_0^\infty J(\omega)/\omega d\omega$.
In spite of this simplicity (\ref{eq:hamilton}) suffices to study decoherence
of the $T_2$-type in the NMR-language which corresponds to 
phase decoherence in the
$XY$-plane of the impurity spin. In this sense, (\ref{eq:hamilton}) constitutes
a minimal, but fully quantum mechanical, model to investigate decoherence 
phenomena.
Spin flips, however, do not occur so that $T_1$ is infinite.
The minimal model renders the analytic examination
of various pulse sequences possible. All thermal, quantum or memory effects 
in the bath as well as back-actions of the qubit on the bath are included.
The pulses used in the following will always be considered
to be ideal, i.e.\ instantaneous  (cf.\ Refs.\
\onlinecite{viola98}) and without any error.

First, we look at a simple measurement assuming 
initially $\sigma_z=1$ and the bath to be in its thermal equilibrium.
Such a state is generated by applying a sufficiently strong
magnetic field in $z$-direction. Then a rotation about
$\sigma_x$ is applied $D_x(\gamma) = \exp(-i\gamma\sigma_x/2) =
\cos(\gamma/2) +i\sigma_x \sin(\gamma/2)$ which transforms
the spin in $z$-direction to
\begin{equation}
\label{eq:flip}
D_x(\gamma)^\dagger \sigma_z D_x(\gamma) = \cos\gamma \sigma_z +
\sin\gamma \sigma_y\ .
\end{equation}
For $\gamma=\pi/2$ a rotation by 90$^\circ$ is achieved; for 
$\gamma=\pi$ the inversion $\sigma^z\to-\sigma^z$. Measuring
$\sigma_y$ leads to  the signal
\begin{eqnarray}
\nonumber
s(t) &=& \langle \uparrow | D_x(\pi/2)^\dagger  \exp(iHt) \sigma_y  
\exp(-iHt)  D_x(\pi/2)| \uparrow \rangle\\
\label{eq:signalb}
 &=& \langle  \uparrow | D_x^\text{ef}(\pi/2)^\dagger
\sigma_y^\text{ef}(t)  
 D_x^\text{ef}(\pi/2) | \uparrow \rangle\ .
\end{eqnarray}
The brackets stand for the thermal expectation value of the bosonic bath.
To obtain the line (\ref{eq:signalb}) 
we transform by $U$ to the effective variables
and define $A(t) := \exp(iHt) A  \exp(-iHt)$. The spin content
of the resulting expression can be calculated using
\begin{subequations}
\label{spinids}
\begin{eqnarray}
\sigma_x^\text{ef}(t) |\uparrow/\downarrow\rangle &=& 
\exp(\mp 2K(t))|\downarrow/\uparrow\rangle\\
\sigma_y^\text{ef}(t) |\uparrow/\downarrow\rangle &=& 
\pm i\exp(\mp 2K(t))|\downarrow/\uparrow\rangle\ .
\end{eqnarray}
\end{subequations}
The bosonic expectation values of exponentials are
computed for operators $A, B$ linear in the bosonic variables with the help of
$\exp(A)\exp(B)=\exp(A+B)\exp([A,B]/2)$ and with the help of
$\langle\exp(A)\rangle =\exp(\langle A^2\rangle/2)$.
In this way, we arrive at
\begin{subequations}
\label{eq:result-simple}
\begin{eqnarray}
s(t) &=& \cos(2\varphi(t)) \exp(-2\chi(t))\quad \text{with}
\\
\varphi(t) &=& \frac{1}{2} 
\int_0^\infty J(\omega) \frac{\sin(\omega t)}{\omega^2} d\omega\\
\label{eq:chi0}
\chi(t) &=&  \int_0^\infty J(\omega) \frac{\sin(\omega t/2)^2}{\omega^2}
\coth(\beta\omega/2) d\omega\ .
\end{eqnarray}
\end{subequations}

\begin{figure}[htbp]
    \begin{center}
     \includegraphics[width=\columnwidth,clip]{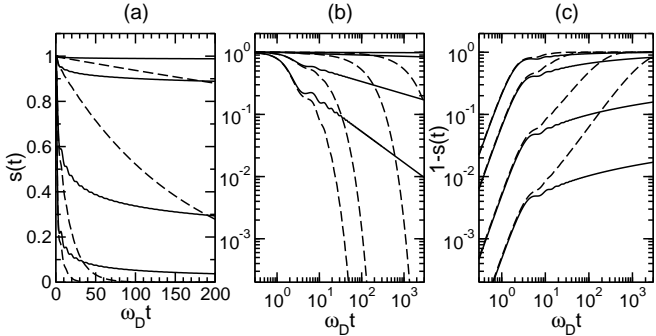}
    \end{center}
    \caption{Signal (\ref{eq:signalb}) vs.\ time. Solid lines for
      $T=0$; dashed ones for $T=0.1\omega_\text{D}$. Panels (a) (linear)
      and (b) (double logarithmic) from bottom to top for $\alpha = 0.25,
      0.1, 0.01, 0.001$. Panel (c) (double logarithmic)
      displays $1-s(t)$ for the same values from top to bottom.}
    \label{fig:unswitched}
\end{figure}
Fig.\ \ref{fig:unswitched} illustrates the effect of
the coupling strength $\alpha$ and of finite temperature $T=1/\beta >0$. 
Panels (a) and (b) display the usual long-time decay while (c) focusses
on the deviation $1-s(t)$ from unity. For quantum
information processes  Fig.\ \ref{fig:unswitched}c shows the relevant data
since $1-s(t)$ should be as low as possible. If error correction
is to be applied thresholds between $10^{-4}$ \cite{alife06,reich06a}
and $10^{-2}$ \cite{knill04,reich04} have to be met. Inspecting
Fig.\ \ref{fig:unswitched}c we conclude that for values of 
$\alpha$  of about $0.1$ the qubit can be stored only for tiny
fractions of the correlation time $t_\text{C}$. But even if
 $\alpha$ is significantly smaller no storage is possible for $t_\text{C}$,
let alone for any time longer.

Another interesting conclusion is that \emph{low} values of
$\omega_\text{D}$ are favorable since they set a \emph{long} time 
scale.\footnote{A numerical estimate yields for 
$\omega_\text{D}=400$K a correlation time of 25 fs which is extremely fast; a 
low value of only $\omega_\text{D}=10$K leads to 1ps which is more favorable.} 
This means that an elastically soft environment, for instance in an organic 
compound with  low $\omega_\text{D}$, is better suited than a hard environment
with high $\omega_\text{D}$. This is counterintuitive, since 
one might have suspected that the influence of vibrations is
lower when the spring constants $\propto \omega_\text{D}^2$ are higher.
The objection that the positive effect of a larger
$t_\text{C}$ in a soft medium will be thwarted by a large value of
the coupling $\alpha$ will be invalidated below.

We pass now to a sequence of pulses where the total
time interval $0\to t$ is split into smaller intervals
$0\to \delta_1 t\to \delta_2 t\to \ldots  \to \delta_n t\to t$
with $0<\delta_1<\delta_2<\ldots < \delta_n < 1$.
The $\delta$-values are taken to be fixed. At each instance $\delta_i t$
a $\pi$-pulse 
$\sigma_y=-iD_y(\gamma=\pi)=-i\exp(-i\gamma\sigma_y/2)|_{\gamma=\pi}$ 
is applied which effectively changes the sign of the interaction in Eq.\ 
(\ref{eq:hamilton}). Hence the observable signal changes from $s(t)$ in
Eq.\ (\ref{eq:signalb}) to
\begin{subequations}
\label{signalc}
\begin{eqnarray}
s_n(t) &=& \langle  \uparrow |D_x^\text{ef}(\pi/2)^\dagger\,
 R^\dagger\, \sigma_y^\text{ef}(t) \, 
 R\,  D_x^\text{ef}(\pi/2) | \uparrow \rangle  \\
R &=& \sigma_y^\text{ef}(\delta_n t)\,
\sigma_y^\text{ef}(\delta_{n-1} t) \ldots \sigma_y^\text{ef}(\delta_2 t)
\,
\sigma_y^\text{ef}(\delta_{1} t) \ . \quad
\end{eqnarray}
\end{subequations}
The evaluation of $s_n(t)$ is based on the same identities as the one of
$s(t)$ except that it is algebraically more involved. The result is
cast in the form
\begin{subequations}
\label{eq:signalc}
\begin{eqnarray}
s_n(t) &=& \cos(2\varphi_n(t)) \exp(-2\chi_n(t))\quad \text{with}
\\
\varphi_n(t) &=&  \int_0^\infty \frac{J(\omega)}{2\omega^2} 
x_n(\omega t)
 d\omega\\
\label{eq:chin}
\chi_n(t) &=&  \int_0^\infty \frac{J(\omega)}{4 \omega^2}
\coth(\beta\omega/2) |y_n(\omega t)|^2 d\omega\ ,
\end{eqnarray}
\end{subequations}
where the factor $x_n(z)$ in the integrand of the phase reads
\begin{equation}
x_n(z) =  (-1)^n\sin(z) +\sum_{m=1}^n(-1)^{m+1}\sin(z\delta_m) 
\end{equation} 
and the factor $y_n(z)$ in the integrand of $\chi_n(t)$ reads 
\begin{equation}
\label{eq:ydef}
y_n(z) =  1 + (-1)^{n+1} e^{iz} + 2\sum_{m=1}^n(-1)^m e^{iz\delta_m} \ .
\end{equation}

The phase is less harmful since its influence on the signal is
only quadratic in $\alpha$. Hence we focus on 
$\chi_n$. The procedure discussed so far is based on $n$ equidistant pulses
\cite{viola98,ban98,cappe06} with $\delta_m = m/(n+1)$ yielding
\begin{subequations}
\begin{eqnarray}
|y_n(z)|^2_\text{eq} &=& 4\tan^2(z/(2n+2))\cos^2(z/2) 
\  \forall n\ \text{even} \qquad\ \\
|y_n(z)|^2_\text{eq} &=& 4\tan^2(z/(2n+2))\sin^2(z/2)  
\ \forall n\ \text{odd} \ .\qquad
\end{eqnarray}
\end{subequations}
For small values of $z/(2n+2)$ these functions rise quadratically like
$(1/2)z^2/(n+1)^2$ omitting rapid oscillations. 
Even without performing the integrations in (\ref{eq:signalc})
one can read off two features: (i) a large number $n$ of pulses is
advantageous. The time scale $t_\text{C}$ is prolonged like
$t_\text{C} \to (n+1) t_\text{C}$. (ii) no further suppression is
achieved since the power law $\propto z^2=(\omega t)^2$ remains
unchanged. 

\begin{figure}[htbp]
    \begin{center}
     \includegraphics[width=\columnwidth,clip]{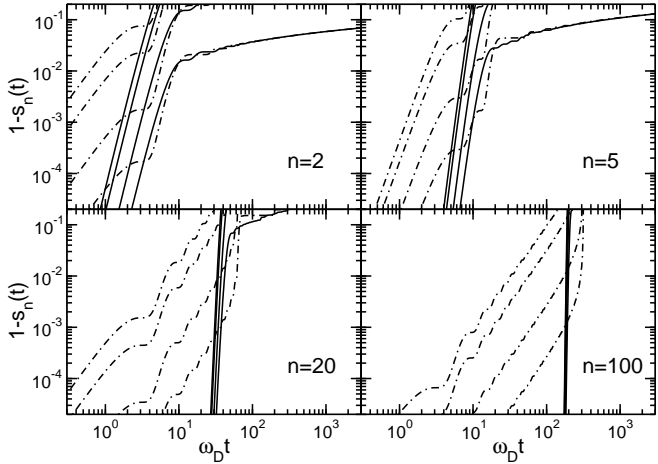}
    \end{center}
    \caption{Signal (\ref{eq:signalc}) vs.\ time for various numbers of pulses
      at $T=0$. Solid lines for the optimized sequence, dashed-dotted lines
      for the equistant sequence (see main text). From top to bottom the curves
      refer to $\alpha = 0.25, 0.1, 0.01, 0.001$. 
    }
    \label{fig:switched}
\end{figure}

Results of the equidistant $\pi$-pulse sequence are shown
in Fig.\ \ref{fig:switched} as dashed-dotted lines. Clearly, a shift
to the right is discernible reflecting the growing factor $n+1$.
But no significant change of the slopes occurs. It is still important
to have a weak coupling between qubit and bosonic bath
to store the qubit for a significant time. For instance
100 pulses make it possible to store the qubit  up to an error of $10^{-4}$
for $\approx 5t_\text{C}$ at $\alpha=0.25$ while for $\alpha=0.001$ it may be 
stored for $\approx 60 t_\text{C}$. 

Now we pose the question whether the sequence of pulses can be optimized, 
for instance by exploiting the freedom of 
choosing the instants of the pulses. The potential of non-equidistant pulse 
sequences was recently demonstrated by concatenated pulse sequences 
\cite{khodj05}. In our work, we aim at finding the \emph{optimum} pulse 
sequence with respect to canonical requirements.

Inspecting (\ref{eq:ydef})
one realizes that $y_n(0)=0$ always and that there are $n$ free parameters
$\{\delta_j\}$. So one may require that $n$ additional conditions
are fulfilled. We use this freedom to make the first $n$ derivatives
$y_n^{(j)}(z)|_{z=0}$ with $j\in\{ 1,2,\ldots n\}$ vanish. 
Nicely, the resulting equations have a simple analytic solution for $n$ 
$\pi$-pulses
\begin{equation}
\label{eq:opt-def}
\delta_j = \sin^2(\pi j/(2n+2))\ .
\end{equation}
This is the main result 
of this Letter. The resulting factor in the integrand yields
\begin{subequations}
\label{eq:integrand-opt}
\begin{eqnarray}
|y_n(z)|^2_\text{op} &=& \left|
\sum\limits_{j=-n-1}^n (-1)^j e^{(iz/2)\cos(\pi j/(n+1))}
\right|^2\\
\label{eq:integrand-opt-b}
 &\approx & 16(n+1)^2 J_{n+1}^2(z/2)
\end{eqnarray}
\end{subequations}
where the second line with the Bessel function $J_{n+1}$
represents a very good approximation valid
for  $z/(2n+2) < 1$ up to exponential corrections. Note that 
$(n+1)J_{n+1}^2(z/2) \propto [z/(2n+2)]^{2n+2}$ 
manifesting the effect of the vanishing leading 
derivatives. From (\ref{eq:integrand-opt}) follows that the
integrand stays extremely small up to a certain value of $z$ of the
order of unity implying that decoherence hardly takes place up to
a certain time $t_\text{op}$ given by  $t_\text{op}\approx (n+1)t_\text{C}$. 
Beyond this time it sets  in very abruptly.

How does our finding compare to known results? For $n=2$ we retrieve from
Eq.~\ref{eq:opt-def} 
$\delta_1=1/4$ and $\delta_2=3/4$. This means that our pulse sequence
with the initial $\pi/2$-pulse about $\sigma_x$ and two $\pi$-pulses
about $\sigma_y$ reproduces the CPMG-cycle \cite{haebe76} which is
widely considered for decoherence suppression \cite{witze06,yao06}.
For all $n>2$, Eq.~\ref{eq:opt-def} predicts so far unexplored pulse
sequences with a better potential for decoherence suppression.

Fig.\ \ref{fig:switched} depicts the features of the optimum sequence.
Clearly, the lines are shifted to the right reflecting the factor
$(n+1)$ in $t_\text{op}$
in parallel to the effect of equidistant pulses. In contrast to
equidistant pulses the optimized pulses lead to steeper and steeper
slopes on increasing $n$ implying that the behavior for different
couplings $\alpha$ becomes almost indistinguishable. This feature is
extremely advantageous because it means practically that even a
large coupling $\alpha$ does not harm a long storage time. For instance
for 100 pulses the qubit may be stored for $\approx 200 t_\text{C}$ 
\emph{independently} of the value of $\alpha$. The possible
storage time is by a factor of 40 better than for the equidistant scheme
 for $\alpha=0.25$. For $\alpha=0.001$ the improvement is still 
about a factor of 4. Clearly, the improvement is most striking for
large values of the coupling. 
Recurring to the estimate of $t_\text{C}=1$ps \cite{endnote23} 
we see that  100 pulses allow us to extend the storage time to about
200ps.

Another way of looking at the optimized scheme Eq.~\ref{eq:opt-def}
is to ask how many pulses are needed to achieve
a certain storage time with an error below a certain threshold, say $10^{-4}$. 
For $\alpha=0.25$,
5 or 6 pulses already imply a storage time of $5t_\text{D}$. For the
same storage the equidistant scheme requires about 100 pulses.
Keeping in mind that in practice each pulse will  be imperfect
 it is certainly advantageous to work with a minimum number of pulses.

Let us turn to temperature.
In practice, no system will be at $T=0$ and in particular
the favorable soft media will be operated at relative high $T$
compared to the cutoff temperature $T_\text{D}=\omega_\text{D}$
(setting $k_\text{B}=1=\hbar$). In our model,
$T=1/\beta$ enters in the
Eqs.\ (\ref{eq:chi0},\ref{eq:chin}) by the $\coth$ factor reflecting
the thermal occupation of the bosonic modes. It deviates from its
$T=0$ value of unity only for small frequencies. Small frequencies
mean small values of $z=\omega t$ so that the suppression of
$|y_n(z)|^2_\text{op}$ in this range, see Eq.\ (\ref{eq:integrand-opt-b}),
is particularly helpful. Finite temperature does not lead
to any noticeable decoherence as long as the storage time is not too long,
i.e.\ as long as it stays below $t_\text{op}$.
Indeed, curves at $T\neq 0$ are indistinguishable from
the solid ones in Fig.\ \ref{fig:switched}. This holds already
for a rather small number of pulses so that it is a highly relevant
feature for experimental realizations. For the equidistant scheme
the thermal effects are larger, in particular for large
temperatures $T\gtrapprox T_\text{D}$. 
The extreme insensitivity to thermal effects
represents another essential advantage of the optimized scheme.

The above derivations hold for arbitrary $T$, i.e.\ even 
for the classical limit  $T\to\infty$. Indeed, the
same pulse sequences can be used for the classical Hamiltonian
$H=(f(t)/2)\sigma_z$ where $f(t)$ is controlled
by  Gaussian fluctuations determined by $\langle f(t)\rangle = 0$
 and by $\langle f(t_1)f(t_2)\rangle =
g(t_1-t_2)$. The Fourier transform $p(\omega)$ of $g(t)$ is the power spectrum
and $p(\omega)/\pi$ replaces $J(\omega)\coth(\beta\omega/2)$ in Eqs.\
(\ref{eq:chi0},\ref{eq:chin}) while the phases $\varphi$ and $\varphi_n$
are zero classically. The other equations remain the same, in particular
Eqs.\ (\ref{eq:opt-def},\ref{eq:integrand-opt}). This observation
greatly increases the applicability of our findings since many systems, not
only bosonic baths, can be described for high temperatures by classical 
Gaussian fluctuations.

The equidistant dynamic decoupling or iterated CPMG cycles
have been realized  experimentally, in particular in NMR experiments.
The detrimental influence of very slow nuclear spins on a solid-state
qubit \cite{frava05} or on the electron
spin in quantum dots  \cite{petta05} has been reduced recently.
A Rabi oscillation could be made vanish by realizing sequences of
almost instantaneous $\pi$-pulses
exploiting the interplay between nuclear and electronic spins 
\cite{morto06}. Krojanski and Suter demonstrated recently that even
the decoherence of large quantum registers, realized by nuclear
spins and their dipole-dipole interaction, can be significantly 
reduced by dynamic decoupling \cite{kroja06}.
To our knowledge, however, no optimized sequences 
obeying Eq.\ (\ref{eq:opt-def}) have been examined. 

An optimized sequence is by definition more efficient than a random
one, cf.~Ref.~\cite{viola05}. But for large symmetry groups
 it may be easier to use
a random scheme than to optimize the pulse sequence. If more
specific information on the bath is available,  cf.~Ref.~\cite{yao06},
other, specifically adapted schemes might work more efficiently.
Furthermore, we emphasize that hybrid techniques are attractive:
Any other dynamic decoupling scheme may be improved by replacing the
$\pi$-pulse or the CPMG cycle of two $\pi$-pulses by a suitable 
$n>2$ sequence obeying Eq.~\ref{eq:opt-def}.
The optimized design of real $\pi$-pulses of finite duration in the
presence of bosonic baths, cf.~Ref.~\cite{motto06} for classical baths, 
is left for future research.

In summary,
we discussed strategies for suppressing the decoherence
of physical quantum bits by dynamic decoupling. A promising way to 
optimize the sequence of $\pi$-pulses beyond the well-known CPMG
sequence was analytically established. 
The comparison to equidistant pulse sequences revealed that
the optimized scheme enhances the possible storage time by up to 
almost two orders of magnitude. Alternatively, the number of pulses
required to achieve a certain prolongation of the storage time 
can be much smaller (by a factor of 20 for strong coupling to the
bosonic bath) than for the standard
equidistant scheme. Additionally, the optimized scheme is extremely
insensitive to  detrimental thermal fluctuations.
So experimental investigations of the optimized scheme are called for.

\acknowledgments

I like to thank F.B.\ Anders, S.\ Pasini, C.\ Raas, J.\ Stolze, and D.\ Suter
for helpful discussions.


\begin{thebibliography}{21}
\expandafter\ifx\csname natexlab\endcsname\relax\def\natexlab#1{#1}\fi
\expandafter\ifx\csname bibnamefont\endcsname\relax
  \def\bibnamefont#1{#1}\fi
\expandafter\ifx\csname bibfnamefont\endcsname\relax
  \def\bibfnamefont#1{#1}\fi
\expandafter\ifx\csname citenamefont\endcsname\relax
  \def\citenamefont#1{#1}\fi
\expandafter\ifx\csname url\endcsname\relax
  \def\url#1{\texttt{#1}}\fi
\expandafter\ifx\csname urlprefix\endcsname\relax\def\urlprefix{URL }\fi
\providecommand{\bibinfo}[2]{#2}
\providecommand{\eprint}[2][]{\url{#2}}

\bibitem[{\citenamefont{Zoller et~al.}(2005)\citenamefont{Zoller, Beth, Binosi,
  Blatt, Briegel, Bruss, Calarco, Cirac, Deutsch, Eisert et~al.}}]{zolle05}
\bibinfo{author}{\bibfnamefont{P.}~\bibnamefont{Zoller}},
  \bibinfo{author}{\bibfnamefont{T.}~\bibnamefont{Beth}},
  \bibinfo{author}{\bibfnamefont{D.}~\bibnamefont{Binosi}},
  \bibinfo{author}{\bibfnamefont{R.}~\bibnamefont{Blatt}},
  \bibinfo{author}{\bibfnamefont{H.}~\bibnamefont{Briegel}},
  \bibinfo{author}{\bibfnamefont{D.}~\bibnamefont{Bruss}},
  \bibinfo{author}{\bibfnamefont{T.}~\bibnamefont{Calarco}},
  \bibinfo{author}{\bibfnamefont{J.~I.} \bibnamefont{Cirac}},
  \bibinfo{author}{\bibfnamefont{D.}~\bibnamefont{Deutsch}},
  \bibinfo{author}{\bibfnamefont{J.}~\bibnamefont{Eisert}},
  \bibnamefont{et~al.}, \bibinfo{journal}{Eur. Phys. J. D}
  \textbf{\bibinfo{volume}{36}}, \bibinfo{pages}{203} (\bibinfo{year}{2005}).

\bibitem[{\citenamefont{Viola and Lloyd}(1998)}]{viola98}
\bibinfo{author}{\bibfnamefont{L.}~\bibnamefont{Viola}} \bibnamefont{and}
  \bibinfo{author}{\bibfnamefont{S.}~\bibnamefont{Lloyd}},
  \bibinfo{journal}{Phys. Rev. A} \textbf{\bibinfo{volume}{58}},
  \bibinfo{pages}{2733} (\bibinfo{year}{1998}).

\bibitem[{\citenamefont{Ban}(1998)}]{ban98}
\bibinfo{author}{\bibfnamefont{M.}~\bibnamefont{Ban}}, \bibinfo{journal}{J.
  Mod. Opt.} \textbf{\bibinfo{volume}{45}}, \bibinfo{pages}{2315}
  (\bibinfo{year}{1998}).

\bibitem[{\citenamefont{Facchi et~al.}(2005)\citenamefont{Facchi, Tasaki,
  Pascazio, Nakazato, Tokuse, and Lidar}}]{facch05}
\bibinfo{author}{\bibfnamefont{P.}~\bibnamefont{Facchi}},
  \bibinfo{author}{\bibfnamefont{S.}~\bibnamefont{Tasaki}},
  \bibinfo{author}{\bibfnamefont{S.}~\bibnamefont{Pascazio}},
  \bibinfo{author}{\bibfnamefont{H.}~\bibnamefont{Nakazato}},
  \bibinfo{author}{\bibfnamefont{A.}~\bibnamefont{Tokuse}}, \bibnamefont{and}
  \bibinfo{author}{\bibfnamefont{D.~A.} \bibnamefont{Lidar}},
  \bibinfo{journal}{Phys. Rev. A} \textbf{\bibinfo{volume}{71}},
  \bibinfo{pages}{022302} (\bibinfo{year}{2005}).

\bibitem[{\citenamefont{Cappellaro et~al.}(2006)\citenamefont{Cappellaro,
  Hodges, Havel, and Cory}}]{cappe06}
\bibinfo{author}{\bibfnamefont{P.}~\bibnamefont{Cappellaro}},
  \bibinfo{author}{\bibfnamefont{J.~S.} \bibnamefont{Hodges}},
  \bibinfo{author}{\bibfnamefont{T.~F.} \bibnamefont{Havel}}, \bibnamefont{and}
  \bibinfo{author}{\bibfnamefont{D.~G.} \bibnamefont{Cory}},
  \bibinfo{journal}{J. Chem. Phys.} \textbf{\bibinfo{volume}{125}},
  \bibinfo{pages}{044514} (\bibinfo{year}{2006}).

\bibitem[{\citenamefont{Haeberlen}(1976)}]{haebe76}
\bibinfo{author}{\bibfnamefont{U.}~\bibnamefont{Haeberlen}},
  \emph{\bibinfo{title}{High Resolution NMR in Solids: Selective Averaging}}
  (\bibinfo{publisher}{Academic Press}, \bibinfo{address}{New York},
  \bibinfo{year}{1976}).

\bibitem[{\citenamefont{Witzel and Sarma}(2006)}]{witze06}
\bibinfo{author}{\bibfnamefont{W.~M.} \bibnamefont{Witzel}} \bibnamefont{and}
  \bibinfo{author}{\bibfnamefont{S.~D.} \bibnamefont{Sarma}},
  \bibinfo{journal}{cond-mat/0604577}  (\bibinfo{year}{2006}).

\bibitem[{\citenamefont{Leggett et~al.}(1987)\citenamefont{Leggett,
  Chakravarty, Dorsey, Fisher, Garg, and Zwerger}}]{legge87}
\bibinfo{author}{\bibfnamefont{A.~J.} \bibnamefont{Leggett}},
  \bibinfo{author}{\bibfnamefont{S.}~\bibnamefont{Chakravarty}},
  \bibinfo{author}{\bibfnamefont{A.~T.} \bibnamefont{Dorsey}},
  \bibinfo{author}{\bibfnamefont{M.~P.~A.} \bibnamefont{Fisher}},
  \bibinfo{author}{\bibfnamefont{A.}~\bibnamefont{Garg}}, \bibnamefont{and}
  \bibinfo{author}{\bibfnamefont{W.}~\bibnamefont{Zwerger}},
  \bibinfo{journal}{Rev. Mod. Phys.} \textbf{\bibinfo{volume}{59}},
  \bibinfo{pages}{1} (\bibinfo{year}{1987}).

\bibitem[{\citenamefont{Weiss}(1999)}]{weiss99}
\bibinfo{author}{\bibfnamefont{U.}~\bibnamefont{Weiss}},
  \emph{\bibinfo{title}{Quantum Dissipative Systems}}
  (\bibinfo{publisher}{World Scientific}, \bibinfo{address}{Singapore},
  \bibinfo{year}{1999}), \bibinfo{edition}{2nd} ed.

\bibitem[{\citenamefont{Aliferis et~al.}(2006)\citenamefont{Aliferis,
  Gottesman, and Preskill}}]{alife06}
\bibinfo{author}{\bibfnamefont{P.}~\bibnamefont{Aliferis}},
  \bibinfo{author}{\bibfnamefont{D.}~\bibnamefont{Gottesman}},
  \bibnamefont{and} \bibinfo{author}{\bibfnamefont{J.}~\bibnamefont{Preskill}},
  \bibinfo{journal}{Quant. Inf. Comput.} \textbf{\bibinfo{volume}{6}},
  \bibinfo{pages}{97} (\bibinfo{year}{2006}).

\bibitem[{\citenamefont{Reichardt}(2006)}]{reich06a}
\bibinfo{author}{\bibfnamefont{B.~W.} \bibnamefont{Reichardt}}, in
  \emph{\bibinfo{booktitle}{Lecture Notes in Computer Science}}
  (\bibinfo{publisher}{Springer}, \bibinfo{address}{Berlin/Heidelberg},
  \bibinfo{year}{2006}), vol. \bibinfo{volume}{4051}, p.~\bibinfo{pages}{50}.

\bibitem[{\citenamefont{Knill}(2004)}]{knill04}
\bibinfo{author}{\bibfnamefont{E.}~\bibnamefont{Knill}},
  \bibinfo{journal}{quant-ph/0404104}  (\bibinfo{year}{2004}).

\bibitem[{\citenamefont{Reichardt}(2004)}]{reich04}
\bibinfo{author}{\bibfnamefont{B.~W.} \bibnamefont{Reichardt}},
  \bibinfo{journal}{quant-phys/0406025}  (\bibinfo{year}{2004}).

\bibitem[{\citenamefont{Khodjasteh and Lidar}(2005)}]{khodj05}
\bibinfo{author}{\bibfnamefont{K.}~\bibnamefont{Khodjasteh}} \bibnamefont{and}
  \bibinfo{author}{\bibfnamefont{D.~A.} \bibnamefont{Lidar}},
  \bibinfo{journal}{Phys. Rev. Lett.} \textbf{\bibinfo{volume}{95}},
  \bibinfo{pages}{180501} (\bibinfo{year}{2005}).

\bibitem[{\citenamefont{Yao et~al.}(2006)\citenamefont{Yao, Liu, and
  Sham}}]{yao06}
\bibinfo{author}{\bibfnamefont{W.}~\bibnamefont{Yao}},
  \bibinfo{author}{\bibfnamefont{R.}~\bibnamefont{Liu}}, \bibnamefont{and}
  \bibinfo{author}{\bibfnamefont{L.~J.} \bibnamefont{Sham}},
  \bibinfo{journal}{cond-mat/0604634}  (\bibinfo{year}{2006}).

\bibitem[{\citenamefont{Fraval et~al.}(2005)\citenamefont{Fraval, Sellars, and
  Longdell}}]{frava05}
\bibinfo{author}{\bibfnamefont{E.}~\bibnamefont{Fraval}},
  \bibinfo{author}{\bibfnamefont{M.~J.} \bibnamefont{Sellars}},
  \bibnamefont{and} \bibinfo{author}{\bibfnamefont{J.~J.}
  \bibnamefont{Longdell}}, \bibinfo{journal}{Phys. Rev. Lett.}
  \textbf{\bibinfo{volume}{95}}, \bibinfo{pages}{030506}
  (\bibinfo{year}{2005}).

\bibitem[{\citenamefont{Petta et~al.}(2005)\citenamefont{Petta, Johnson,
  Taylor, Laird, Yacoby, Lukin, Markus, Hanson, and Gossard}}]{petta05}
\bibinfo{author}{\bibfnamefont{J.~R.} \bibnamefont{Petta}},
  \bibinfo{author}{\bibfnamefont{A.~C.} \bibnamefont{Johnson}},
  \bibinfo{author}{\bibfnamefont{J.~M.} \bibnamefont{Taylor}},
  \bibinfo{author}{\bibfnamefont{E.~A.} \bibnamefont{Laird}},
  \bibinfo{author}{\bibfnamefont{A.}~\bibnamefont{Yacoby}},
  \bibinfo{author}{\bibfnamefont{M.~D.} \bibnamefont{Lukin}},
  \bibinfo{author}{\bibfnamefont{C.~M.} \bibnamefont{Markus}},
  \bibinfo{author}{\bibfnamefont{M.~P.} \bibnamefont{Hanson}},
  \bibnamefont{and} \bibinfo{author}{\bibfnamefont{A.~C.}
  \bibnamefont{Gossard}}, \bibinfo{journal}{Science}
  \textbf{\bibinfo{volume}{309}}, \bibinfo{pages}{2180} (\bibinfo{year}{2005}).

\bibitem[{\citenamefont{Morton et~al.}(2006)\citenamefont{Morton, Tyryshkin,
  Ardavan, Benjamin, Porfyrakis, Lyon, and Briggs}}]{morto06}
\bibinfo{author}{\bibfnamefont{J.~J.~L.} \bibnamefont{Morton}},
  \bibinfo{author}{\bibfnamefont{A.~M.} \bibnamefont{Tyryshkin}},
  \bibinfo{author}{\bibfnamefont{A.}~\bibnamefont{Ardavan}},
  \bibinfo{author}{\bibfnamefont{S.~C.} \bibnamefont{Benjamin}},
  \bibinfo{author}{\bibfnamefont{K.}~\bibnamefont{Porfyrakis}},
  \bibinfo{author}{\bibfnamefont{S.~A.} \bibnamefont{Lyon}}, \bibnamefont{and}
  \bibinfo{author}{\bibfnamefont{G.~A.~D.} \bibnamefont{Briggs}},
  \bibinfo{journal}{Nature Phys.} \textbf{\bibinfo{volume}{2}},
  \bibinfo{pages}{40} (\bibinfo{year}{2006}).

\bibitem[{\citenamefont{Krojanski and Suter}(2006)}]{kroja06}
\bibinfo{author}{\bibfnamefont{H.~G.} \bibnamefont{Krojanski}}
  \bibnamefont{and} \bibinfo{author}{\bibfnamefont{D.}~\bibnamefont{Suter}},
  \bibinfo{journal}{Phys. Rev. Lett.} \textbf{\bibinfo{volume}{97}},
  \bibinfo{pages}{150503} (\bibinfo{year}{2006}).

\bibitem[{\citenamefont{Viola and Knill}(2005)}]{viola05}
\bibinfo{author}{\bibfnamefont{L.}~\bibnamefont{Viola}} \bibnamefont{and}
  \bibinfo{author}{\bibfnamefont{E.}~\bibnamefont{Knill}},
  \bibinfo{journal}{Phys. Rev. Lett.} \textbf{\bibinfo{volume}{94}},
  \bibinfo{pages}{060502} (\bibinfo{year}{2005}).

\bibitem[{\citenamefont{M\"ott\"onen et~al.}(2006)\citenamefont{M\"ott\"onen,
  de~Sousa, Zhang, and Whaley}}]{motto06}
\bibinfo{author}{\bibfnamefont{M.}~\bibnamefont{M\"ott\"onen}},
  \bibinfo{author}{\bibfnamefont{R.}~\bibnamefont{de~Sousa}},
  \bibinfo{author}{\bibfnamefont{J.}~\bibnamefont{Zhang}}, \bibnamefont{and}
  \bibinfo{author}{\bibfnamefont{K.~B.} \bibnamefont{Whaley}},
  \bibinfo{journal}{Phys. Rev. A} \textbf{\bibinfo{volume}{73}},
  \bibinfo{pages}{022332} (\bibinfo{year}{2006}).

\end{thebibliography}

\end{document}